\documentclass{sig-alternate}
\usepackage{verbatim}
\usepackage{url}

\newcommand\T{\rule{0pt}{2.6ex}}
\newcommand\B{\rule[-1.2ex]{0pt}{0pt}}

\begin{document}

 \conferenceinfo{M-MPAC'2011,}{December 12th, 2011, Lisbon, Portugal.}
 \CopyrightYear{2011}
 \crdata{978-1-4503-1065-9/11/12}

\title{Instance-based XML Data Binding for Mobile Devices}
\numberofauthors{3}

\author{
% 1st. author
\alignauthor
Alain Tamayo\\
       \affaddr{Institute of New Imaging Technologies}\\
       \affaddr{Universitat Jaume I, Spain}\\
       \affaddr{Ave Vicent Sos Baynat, SN, 12071, Castell\'on de la Plana}\\
       \affaddr{atamayo@uji.es}\\
       %\email{atamayo@uji.es}
% 2nd. author
\alignauthor
Carlos Granell\\
      \affaddr{Institute for Environment \\and Sustainability \\European Commission}\\
       \affaddr{Joint Research Centre}\\
       \affaddr{Ispra, Italy}\\
       \affaddr{carlos.granell@jrc.ec.europa.eu}\\
       %\email{carlos.granell@jrc.ec.europa.eu}
% 3rd. author
\alignauthor 
Joaqu\'in Huerta\\
      \affaddr{Institute of New Imaging Technologies}\\
       \affaddr{Universitat Jaume I, Spain}\\
       \affaddr{Ave Vicent Sos Baynat, SN, 12071, Castell\'on de la Plana}\\
       \affaddr{huerta@uji.es}\\
       %\email{huerta@uji.es}
}

\maketitle

\begin{abstract}

XML and XML Schema are widely used in different domains for the definition of standards that enhance the interoperability between parts exchanging information through the Internet. 
The size and complexity of some standards, and their associated schemas, have been growing with time as new use case scenarios and data models are added to them. 
The common approach to deal with the complexity of producing XML processing code based on these schemas is the use of XML data binding generators. 
Unfortunately, these tools do not always produce code that fits the limitations of resource-constrained devices, such as mobile phones, in the presence of large schemas. 
In this paper we present \textit{Instance-based XML data binding}, an approach to produce compact application-specific XML processing code for mobile devices. 
The approach utilises information extracted from a set of XML documents about how the application make use of the schemas. 
\end{abstract}



 % NOT required for Proceedings

\section{Introduction}

eXtensible Markup Language (XML) has reached a great success in the Internet era. 
XML documents are similar to HTML documents, but do not restrict users to a single vocabulary, which offers a great deal of flexibility to represent information. 
To define the structure of documents within a certain vocabulary, schema languages such as \textit{Document Type Definition} (DTD) or \textit{XML Schema} are used. 

XML has been adopted as the most common form of encoding information exchanged by Web services  \cite{proc:kay, article:wilde2, article:wilde}. 
\cite{proc:kay} attribute this success to two reasons. 
The first one is that the XML specification is accessible to everyone and it is reasonably simple to read and understand. 
The second one is that several tools for processing XML are readily available. 
We add to these reasons that as XML is \textit{vocabulary-agnostic}, it can be used to represent data in basically any domain. 
For example, we can find the \textit{Universal Business Language} (UBL)\footnote{http://docs.oasis-open.org/ubl/cs-UBL-2.0/UBL-2.0.html} in the business domain, or the standards defined by the \textit{Open Geospatial Consortium} (OGC) in the geospatial domain. 
UBL defines a standard way to represent business documents such as electronic invoices or  electronic purchase orders. OGC standards define \textit{web service interfaces} and \textit{data encodings} to exchange geospatial information. 
All of these standards (UBL and OGC's) have two things in common. 
The first one is that they use XML Schema to define the structure of XML documents. 
The second one is that the size and complexity of the standards is very high, making very difficult its manipulation or implementation in certain scenarios \cite{proc:pichler, proc:tamayo3}.  

The use of such large schemas can be a problem when XML processing code based on the schemas is produced for a resource-constrained device, such as a mobile phone. 
This code can be produced using a manual approach, which will require the low-level manipulation of XML data, often producing code that is hard to modify and maintain. 
Another option is to use an XML data binding code generator that maps XML data into application-specific concepts. 
This way developers can focus on the semantics of the data they are manipulating \cite{proc:white}. 
The problem with generators is that they usually make a straightforward mapping of schema components to programming languages constructs that may result in a binary code with a very large size that cannot be easily accommodated in a mobile device \cite{proc:tamayo3}.  

Although schemas in a certain domain can be very large this does not imply that all of the information contained on them is necessary for all of the applications in the domain. 
For example, in \cite{coll:tamayo1} a study of the use of XML in a group of 56 servers implementing the \textit{OGC's Sensor Observation Service (SOS) specification}\footnote{SOS is a standard web service interface to exchange information between sensor data producers and consumers\cite{ogc:sos}.}  revealed that only 29.2\% of the SOS schemas were used in a large collection of XML documents gathered from those servers. 
Based on this information we proposed in \cite{proc:tamayo4} an algorithm to simplify large XML schema sets in an application-specific manner by using a set of XML documents conforming to these schemas. 
The algorithm allowed a 90\% reduction of the size of the schemas for a real case study. 
This reduction was translated in a reduction of binary code ranging between 37 to 84\% when using code generators such as JAXB\footnote{https://jaxb.dev.java.net}, XMLBeans\footnote{http://xmlbeans.apache.org} and XBinder\footnote{http://www.obj-sys.com/xbinder.shtml}.

In this paper we extend the schema simplification algorithm presented in \cite{proc:tamayo4} to a more complete \textit{Instance-based XML Data Binding} approach. 
This approach allows to produce very compact application-specific XML processing code for mobile devices. 
In order to make the code as small as possible the approach will use, similarly to \cite{proc:tamayo4}, a set of XML documents conforming to the application schemas. 
From these documents, in addition to extract the subset of the schemas that is needed, we extract other relevant information about the use of schemas that can be utilised to reduce the size of the final code.  
A prototype implementation targeted to Android\footnote{http://www.android.com} and the Java programming language has been developed.  

The remainder of this paper is structured as follows. Section 2 presents an introduction to XML Schema and XML data binding. In Section 3, related work is presented. The \textit{Instance-based data binding approach} is presented in Section 4. Section 5 overviews some implementation details and limitations found during the development of the prototype. Section 6 presents experiments to measure size an execution times of the code generated by the tool in a real scenario. Last, conclusions and future work are presented. 

\section{Background}

In this section we present a brief introduction to the topics of XML Schema and XML data binding. XML Schema files are used to assess the validity of well-formed element and attribute information items contained in XML instance files \cite{w3c:schemas1}\cite{w3c:schemas2}. The term XML data binding refers to the idea of taking the information in an XML document and convert it to instances of application objects \cite{book:mclaughlin}.

\subsection{XML Schema}

An XML Schema document contains components in the form of complex and simple type definitions, element declarations, attribute declarations, group definitions, and attribute group definitions. This language allows users to define their own types, in addition to a set of predefined types defined by the language. Elements are used to define the content of types and when global, to define which of them are valid as top-level element of an XML document.

XML Schema provides a derivation mechanism to express subtyping relationships. This mechanism allows types to be defined as subtypes of existing types, either by extending or restricting the content model of base types . Apart from type derivation, a second subtyping mechanism is provided through substitution groups. This feature allows global elements to be substituted by other elements in instance files. A global element E, referred to as \textit{head element}, can be substituted by any other global element that is defined to belong to the E's substitution group.

\subsection{XML Data Binding}

With \textit{XML Data Binding}, an abstraction layer is added over the raw XML processing code, where XML information is mapped to data structures in an application data model. XML data binding code is often produced by using code generators that use a description of the structure of XML documents using some schema language. The use of generators potentially gives benefits such as increased productivity, consistent quality throughout all the generated code, higher levels of abstraction as we usually work with an abstract model of the system; and the potential to support different programming languages, frameworks and platforms \cite{book:herrington}. 

\begin{sloppypar}
Although most of the generators available nowadays are targeted to desktop or server applications, several tools have been develop for mobile devices such as XBinder and CodeSysnthesis XSD/e\footnote{http://codesynthesis.com/products/xsde}, or for building complete web services communication end-points for resource constrained environments, such as gSOAP \cite{proc:vanengelen}. All of the tools mentioned before map XML Schema structures to programming languages construct in a straightforward way, which is not adequate when large schemas sets are used.
\end{sloppypar}

\section{Related Work}

\begin{sloppypar}
Problems related with having large and complex schemas have been presented in several articles \cite{proc:pichler, coll:rahm, proc:tamayo4,  proc:villegas}. 
For example, \cite{proc:pichler} deal with problems of large schemas in schema matching in the business domain. 
In the context of schema and ontology mapping, \cite{coll:rahm} states that current match systems still struggle to deal with large-scale match tasks to achieve both good effectiveness and good efficiency.
\cite{proc:tamayo4}, the work extended here, expose the problems related to using XML data binding tools to generate XML processing code for mobile geospatial applications.
Last, \cite{proc:villegas} present an algorithm to extract fragments of large conceptual schemas arguing that the largeness of these schemas makes difficult the process of getting the knowledge of interest to users.   
\end{sloppypar}

When considering XML processing in the context of mobile devices, literature is focused in two main competing requirements: \textit{compactness} (of information) and \textit{processing efficiency} \cite{article:kangasharju1}. 
To achieve compactness compression techniques are used to reduce the size of XML-encoded information \cite{proc:kabisch, proc:kangasharju2, w3c:exi}. 
About processing efficiency, not much work has been done in the mobile devices field. 
A prominent exception in this topic is the work presented in \cite{article:kangasharju1}, \cite{proc:kangasharju2} and \cite{proc:lindholm}. 
These articles are all related to the implementation of a middleware platform for mobile devices: the \textit{Fuego mobility middleware} \cite{proc:tarkoma}, where XML processing has a large impact. 
The proposed \textit{XML stack} provides a general-purpose XML processing API called \textit{XAS} \cite{article:kangasharju1}, an XML binary format called \textit{Xebu} \cite{proc:kangasharju2}, and others APIs such as \textit{Trees-with-references} (RefTrees) and \textit{Random Access XML Store} (RAXS)\cite{proc:lindholm}. 

Regarding the use of instance files to drive the manipulation of schemas, \cite{coll:rahm} presents a review of different methods that use instance files for ontology matching. 
In the field of  schema inference, instance files are used as well to generated adequate schema files that can be used to assess their validity (e.g. \cite{proc:bex, proc:hegewald, proc:min}).

\section{Instance-based Data Binding}  

\textit{Instance-based XML data binding}, is a two-step process. 
The first step, \textit{Instance-based schema simplification}, extracts the information about how schema components are used by a specific application, based on the assumption that a representative subset of XML documents that must be manipulated by the application is available.  The second step, \textit{Code generation}, consists of using all of the information extracted in the previous step to generate XML processing code as optimised as possible for a target platform.

\begin{figure}
  % Requires \usepackage{graphicx}
  \centering
  \includegraphics[scale=0.6]{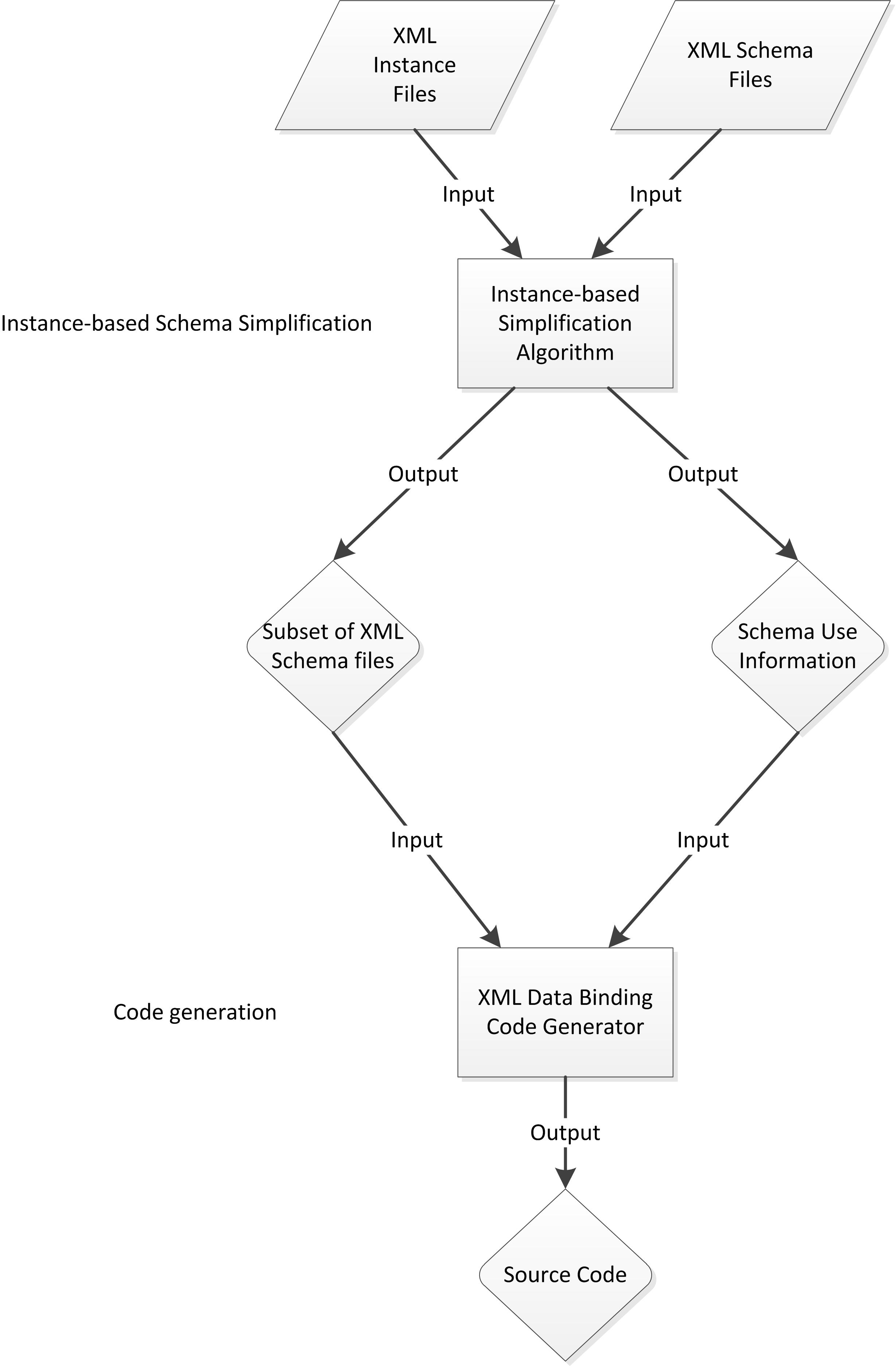}\\
  \caption{Instance-based XML data binding code generation process}\label{fig:flow-XMLdatabinding}
\end{figure}

The whole process is shown in Figure \ref{fig:flow-XMLdatabinding}, the inputs to the first step are a set of schemas and a set of XML documents conforming to them. The outputs will be the subset of the schemas used by the XML documents and other information about the use of certain features of the schemas that can be used to optimise the code in the following step. The outputs of the first step are the inputs of the code generation step. The two steps of the process are detailed in the following subsections.

\subsection{Instance-based Schema Simplification}  

The \textit{Instance-based schema simplification} step extracts the subset of the schemas used on a set of XML documents. The algorithm used to perform this simplification was first presented in \cite{proc:tamayo4} and has been extended here to extracts other information that can be used to produce  more compact XML processing code. 

\begin{figure}
  % Requires \usepackage{graphicx}
  \centering
  \includegraphics[scale=0.034]{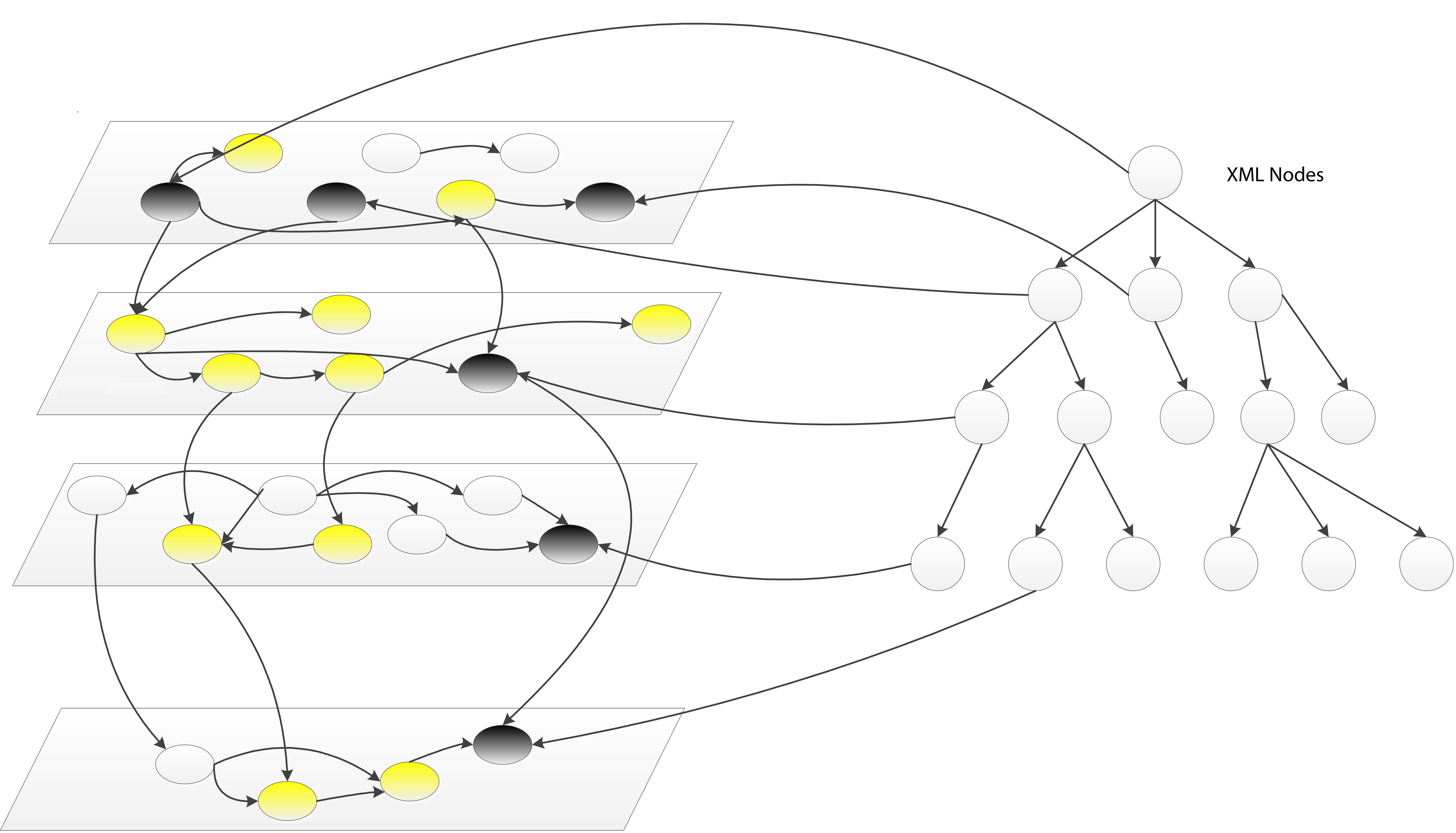}\\
  \caption{Relations between XML instance information items (right) and schemas components defining its structure (left)}\label{fig:simplificationAlg}
\end{figure}

The idea behind this algorithm, is depicted graphically in Figure \ref{fig:simplificationAlg}. 
The figure shows to the left the graph of relationships between schemas components. 
The different planes represent different namespaces. Links between schema components represent dependencies between them. 
To the right we have the tree of information items (XML nodes) contained in XML documents.
For the sake of simplicity we show in the figure only the tree of nodes corresponding to a single document. 
An edge between an XML node and a schema component represents that the component describes the structure of the node. 
To simplify the figure we have shown only a few edges, although an edge for every XML node must exist.  
Starting from a set of XML documents and the schema files defining their structure, it is possible to calculate which schema components are used and which are not. 
In doing so, the following information is also recorded:

\begin{itemize}
\item \textit{Types that are instanced in XML documents}: For each XML node exists a schema type describing its structure. While XML documents are processed the type of each XML node is recorded. This way we can know which types are instanced and which are not.
\item \textit{Types and elements substitutions}: The subtyping mechanisms mentioned in Section 2.1 allow  the \textit{real} or \textit{dynamic type} of an element to be different from its \textit{declared type}. Elements declared as having type A, may have any type derived from A in an XML document. In this case the real type must be specified with the attribute \textit{xsi:type}. Something similar happens with substitution groups, although in this case the attribute \textit{xsi:type} is not necessary. The information about XML nodes whose dynamic type is different from its declared type is recorded.  
\item \textit{Wildcards substitutions}: The elements used to substitute wildcards are recorded. 
\item \textit{Elements occurrence constraints information}: For all of the elements it is checked that if they allow multiple occurrences there is at least one document where several occurrences of the element are present.
\item \textit{Elements with a single child}: All of the elements that contain a single child  are also recorded.
\end{itemize}

\begin{sloppypar}

\end{sloppypar}

\subsection{Code generation process}

\begin{figure}
\centering
  \includegraphics[scale=0.64]{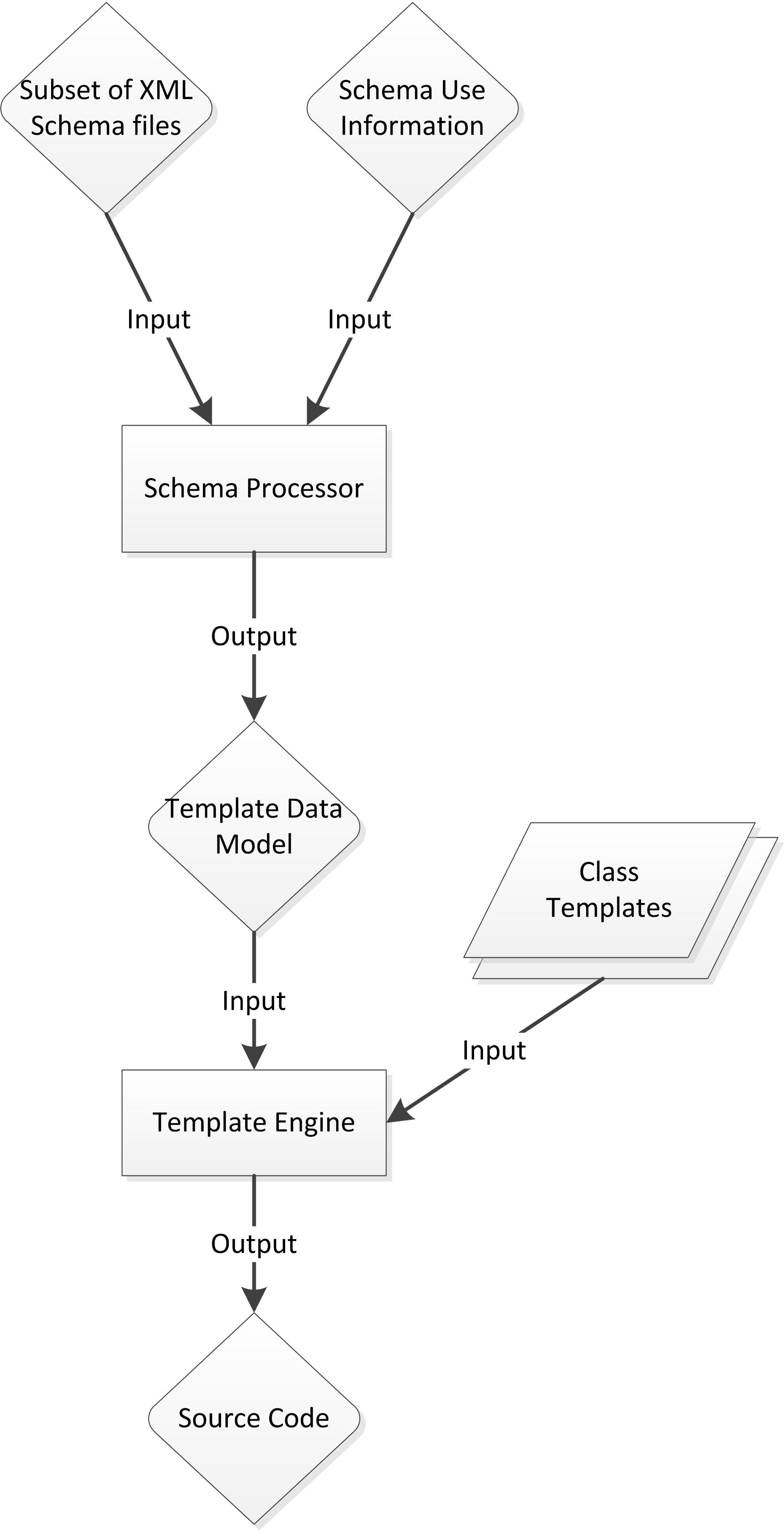}\\
  \caption{Flow diagram for the code generation process}\label{fig:flow_gen}
\end{figure}

A more detailed view of the code generation process is shown in Figure \ref{fig:flow_gen}. The outputs of the schema simplification step are used as inputs to the \textit{schema processor}, the component of the generator in charge of creating the data model that will be used later by the \textit{template engine}. The \textit{template engine} combines pre-existing \textit{class templates} with the data model to generate the final source code. The use of a template engine allows the generation of code for other platforms and programming languages by just defining new class templates.

%\subsubsection{Supported Features}

A summary of the features of the code generation process that contribute to the generation of optimised code is listed next: 

\begin{sloppypar}
\begin{itemize}
\item \textit{Use of information extracted from XML documents}: The use of  information about schema use allows to apply the following optimisations:  			 
\begin{itemize}
			\item \textit{Remove unused schema components}: The schema components that are not used are not considered for code generation. By removing the unused components we can substantially reduce the size of the generated code. The amount of the reduction will depend on how specific applications make use of the original schemas. 
			\item \textit{Efficient handling of subtyping and wildcards}: The number of possible substitutions of a type by its subtypes, and a head element of a substitution group by the members of the group can be bounded with the information gathered from the instances files. In the general case, where no instance-based information is available, generic code to face any possible type or element substitution must be written. Limiting the number of possible substitutions to only a few allows the production of simpler and faster code. The same reasoning is applied to wildcards.
			\item \textit{Inheritance flattening}: By flattening subtyping hierarchies for a given type, i.e., including explicitly in its type definition all of the fields inherited from base types and eliminating the subtype relationship with its parents, we can reduce the number of classes in the generated code.  The application of this technique  will not necessarily result in smaller generated code, as the fields defined in base types must be replicated in all of their child types, but it will have a positive impact in the work of the class loader because a lower number of classes have to be loaded while the application is executed. Let us consider the case of the geospatial schemas introduced in Section 1. These schemas typically present deep subtyping hierarchies with six or more levels, as a consequence when an XML node of a type in the lowest levels of the hierarchy must be processed, all of its parent types must be loaded first. The technique of inheritance flattening has been widely explored and used in different computer science and engineering fields as is proven by the abundant literature found in the topic \cite{proc:beyer, proc:ciprian, proc:bungartz,  article:bungartz, proc:ciccheti, proc:lagorio}.
			\item \textit{Adjust occurrence constraints}: If an element is declared to have multiple occurrences it must be mapped to a data structure in the target programming language that allows the storage of the multiple instances of the elements, e.g. an array or a linked list. In practice if the element has at most one occurrence in the XML documents that must be processed by the application it can be mapped to a single object instance. Using this optimisation the final code will make a better use of memory because  instead of creating a collection (array, linked list, etc.) that will only contain a single object, it  creates a single object instance.
\end{itemize}
\item \textit{Collapse elements containing single child elements}: Information items that will always contain single elements can be replaced directly by its content. By applying this optimisation we can reduce the number of classes in the generated code, which will have a positive impact in the size of the final code, the amount of work that has to be done by the class loader, and the use of memory during execution. This optimization is used by mainstream XML data binding tools such as JiBX\footnote{http://jibx.sourceforge.net/} and the XML Schema Definition Tool\footnote{http://msdn.microsoft.com/en-us/library/x6c1kb0s}.
\item \textit{Disabling parsing/serialization operations as needed}: Some code generators always includes code for parsing and serialization even when only one of these functions is needed. For example, in the context of geospatial web services, most of the time spent in XML processing by client-side applications is dedicated to parsing, as messages received from the servers are potentially large. On the other hand,  most of these services allows request to be sent to the server encoded in an HTTP GET request, therefore XML serialisation is not needed at all.
\item \textit{Ignoring sections of XML documents}: Frequently, we are not interested in all of the information contained in XML files, ignoring the unneeded portions of the file will improve the speed of the parsing process and it may have a significant impact in the amount of memory used by the application.
\end{itemize}
\end{sloppypar}

In addition, the following features not related directly with code optimisation are also supported:

\begin{itemize}
\item \textit{Source code based on simple code patterns}: The generated source code is straightforward to understand and modify in case it is necessary.
\item \textit{Tolerate common validation errors:} Occasionally,  XML documents that are not valid against their respective schemas must be processed by our applications. In many cases, the validation errors can be ignored following simple coding rules.
\end{itemize}

A detailed explanation of each of the features presented in this section can be found in \cite{thesis:tamayo}.

As mentioned before, the approach presented in this paper is based on the assumption that a representative set of XML documents exists. By \textit{representative} we mean that these documents contain instances of all of the possible XML Schema elements and types that will be processed by the application in the future. Nevertheless, this subset might not always be available. In this case, we can still take advantage of the approach by building \textit{synthetic} XML documents containing relevant information. Whether XML processing code is produced manually or automatically developers typically have some knowledge of the structure of the documents that must be processed by the applications. Therefore, we can use this knowledge to build  sample XML documents that can be used as input to the algorithm. In case it were necessary, the final code can be manually modified later, or the sample files changed and used to regenerate the code.

If we were using synthetic documents instead of actual documents some of the optimisations related to the information extracted from them should not be applied. The reason for this is that  we do not have enough information about how the related schema features are used. For example, we cannot apply optimisations such as the efficient handling of subtyping and wildcards, as we might not know all of the possible type substitutions. Something similar happens with the adjustment of occurrence constraints. Nevertheless, other optimisations such as inheritance flattening or removing  unused schema components can be still safely applied.

\section{Implementation}

\textit{DBMobileGen} (DBMG for short) is the current implementation of the \textit{Instance-based XML data binding} approach \cite{thesis:tamayo}. It includes components implementing both the simplification algorithm and code generation process. It is implemented in Java and relies on existing libraries such as \textit{Eclipse XSD}\footnote{http://www.eclipse.org/modeling/mdt/?project=xsd\#xsd} for processing XML schemas, \textit{Freemarker}\footnote{http://freemarker.sourceforge.net} as template engine library, and as well as the generated code, \textit{kXML}\footnote{http://kxml.sourceforge.net/kxml2/} for low-level XML processing. This tool produces code targeted to Android mobile devices and the Java programming language.

The current implementation has some limitations. Because of the complexity of the XML Schema language itself, support for certain features and operations have been only included if it is considered necessary for the case study or applications where the tool has been used \cite{pre:arago, proc:tamayo5}. Some of these limitations are  listed next:

\begin{itemize}
\item \textit{Serialization is not supported yet}: The role of parsing for our sample applications and case studies is far more important than serialization. This is mainly because we have preferred to use HTTP GET to issue server requests wherever possible. 
\item  \textit{Dynamic typing using xsi:type not fully supported}: The mechanism of dynamic type substitution by using the \textit{xsi:type} attribute has not been fully implemented yet, as the XML documents processed in the applications developed so far do not use this feature. 
\end{itemize}

\section{Experiments}

In this section we present two experiments. The first one tries to test how much the size of the generated code can be reduced by using DBMG. The second one measures the execution times of generated code in a mobile phone.

\subsection{Measuring code size}

In this experiment we borrow the test case presented in \cite{proc:tamayo4} that implements the communication layer for an SOS client. SOS is a standard web service interface defined to enhance interoperability between sensor data producers and consumers \cite{ogc:sos}.  The SOS schemas are among the most complex geospatial web service schemas as they are comprised of more than 80 files and they contain more than 700 complex types and global elements \cite{proc:tamayo3} .

The client must process data retrieved  from a server that contains information about air quality for the Valencian Community. This information is gathered by  57 control stations located in that area. The stations measure the level of different contaminants in the atmosphere.

A set of 2492 XML documents was gathered from the server to be used as input, along with the SOS schemas, to the Instance-based data binding process. The source code generated by DBMG is compiled to the \textit{compressed jar} format and compared with the final code generated by other generators: XBinder, JAXB and XMLBeans. The last two are not targeted to mobile devices but are used here as reference to compare the size of similar code for other types of applications.

Table \ref{generatedcode2} shows the size of the code produced with the different generators from the full SOS schemas (Full) and from the subset of the schemas used in the input instance files (Reduced). The reduced schemas are calculated applying the schema simplification algorithm to the full SOS schemas. The last row of the table (Libs) includes the size of the supporting libraries needed to execute the generated code in each case. 

\begin{table}
\centering
\caption{Comparing size of code (KBs) for original and simplified schema sets}\label{generatedcode2}
\begin{tabular}{ |p{1.5cm}| p{1.1cm}| p{1cm}| p{1.6cm}| p{1cm}|} \hline
 \T \B &XBinder & JAXB & XMLBeans & DBMG\\ \hline
Full         & 3,626 &   754 & 2,822 & 88\\\hline
Reduced      &   567 &    90 &   972 & 88\\\hline
Libs 		 &   100 & 1,056 & 2,684 & 30\\\hline
\end{tabular}
\end{table}

\begin{figure}
 \begin{center}
\includegraphics[scale=0.20]{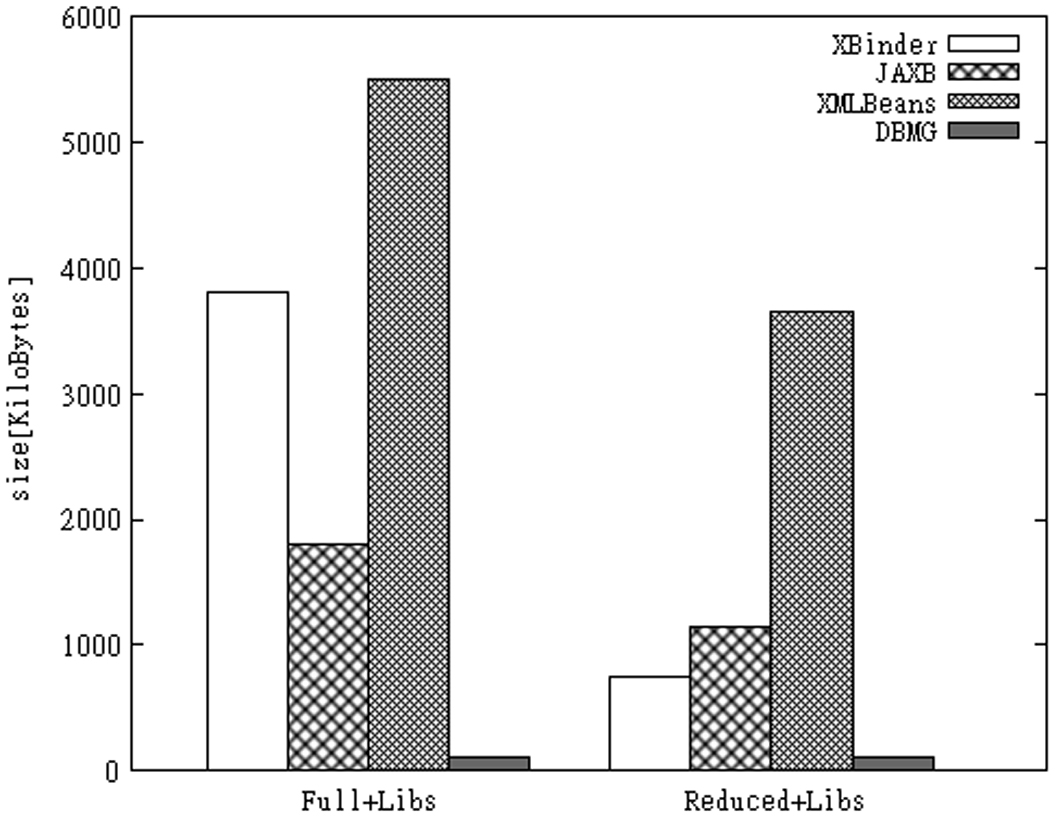}\\
\caption{Size of generated code for full schemas}\label{fig:codegen_full}
\end{center}
\end{figure}

Figure \ref{fig:codegen_full} shows the total size of XML processing code when using the full and reduced schemas. In both cases, we can see the enormous difference that exists between the code generated by DBMG and the code generated by other tools. 

The size for DBMG is the same in both cases because it implicitly performs the simplification of the schemas before generating source code. It must be noted that serialisation is not still implemented in DBMG. We roughly estimate that including serialisation code will increase the final  size in about 30\%. In any case, the code generated by this tool is about 6 times smaller than the code generated by XBinder from the reduced schemas and about 30 times smaller than the code generated from the full schemas. One of the reasons for this difference in size is the lack of serialisation support in DBMG. Another reason is that XBinder generates code to ensure all of the restrictions related to user-defined simple types. This is an advantage if we parse data obtained for a non-trusted source and the application requires the data to be carefully validated, but it is a disadvantage in the opposite case, as unneeded verification increase processor usage and memory footprint. In the case of DBMG, as it aggressively tries to lower final code size, these simple type restrictions are ignored and these types do not even have a counterpart in the generated code. 

When compared to JAXB, using the reduced schemas, the main difference in size is in the supporting libraries, as the code generated by JAXB is very simple. Still, the code generated by DBMG is slightly smaller because the step of removing elements with single child elements and inheritance flattening eliminates a large number of classes.  In all of the cases, XMLBeans has the largest size. This tool is mostly optimised for speed at the expense of generating a more sophisticated and complex code and the use of bigger supporting libraries.  

\subsection{Measuring execution times}
 
To test the performance of the generated code  we will parse a set of 38 capabilities files\footnote{Capabilities files contain metadata about the information contained in a server instance that implement any of the  OGC's web service interfaces.} obtained from different SOS servers. The code needed to parse these files is generated and deployed to a HTC Desire Android smartphone with a 1 GHz Qualcomm QuadDragon CPU and 576 MB of RAM. The 38 files have sizes ranging from less than 4 KB to 3.5 MB, with a mean size of 315 KB and a standard deviation of 26.7 KB. As the size range is large and with the purpose of simplifying presentation we divide the files in two groups, those with a size below 100 KB, CAPS-S (30 files), and those with size equal to or higher than 100 KB, CAPS-L (8 files).

To obtain accurate measures of the execution time for the code we selected the methodology presented in \cite{proc:georges}. This methodology provides a statistically rigorous approach to deal with all of the non-deterministic factors that may affect the measurement process (multi-threading, background processes, etc.).

As our goal is only to measure the execution times of XML processing code, we stored the files to be parsed locally  to avoid interferences related to network delays. Besides, to minimise the interference of data transfer delays from the storage medium all of the files below 500 KB were read into memory before being parsed. It was impossible to do the same for files with sizes above 500 KB because of the device memory restrictions.

\begin{figure}
 \begin{center}
\includegraphics[scale=0.12]{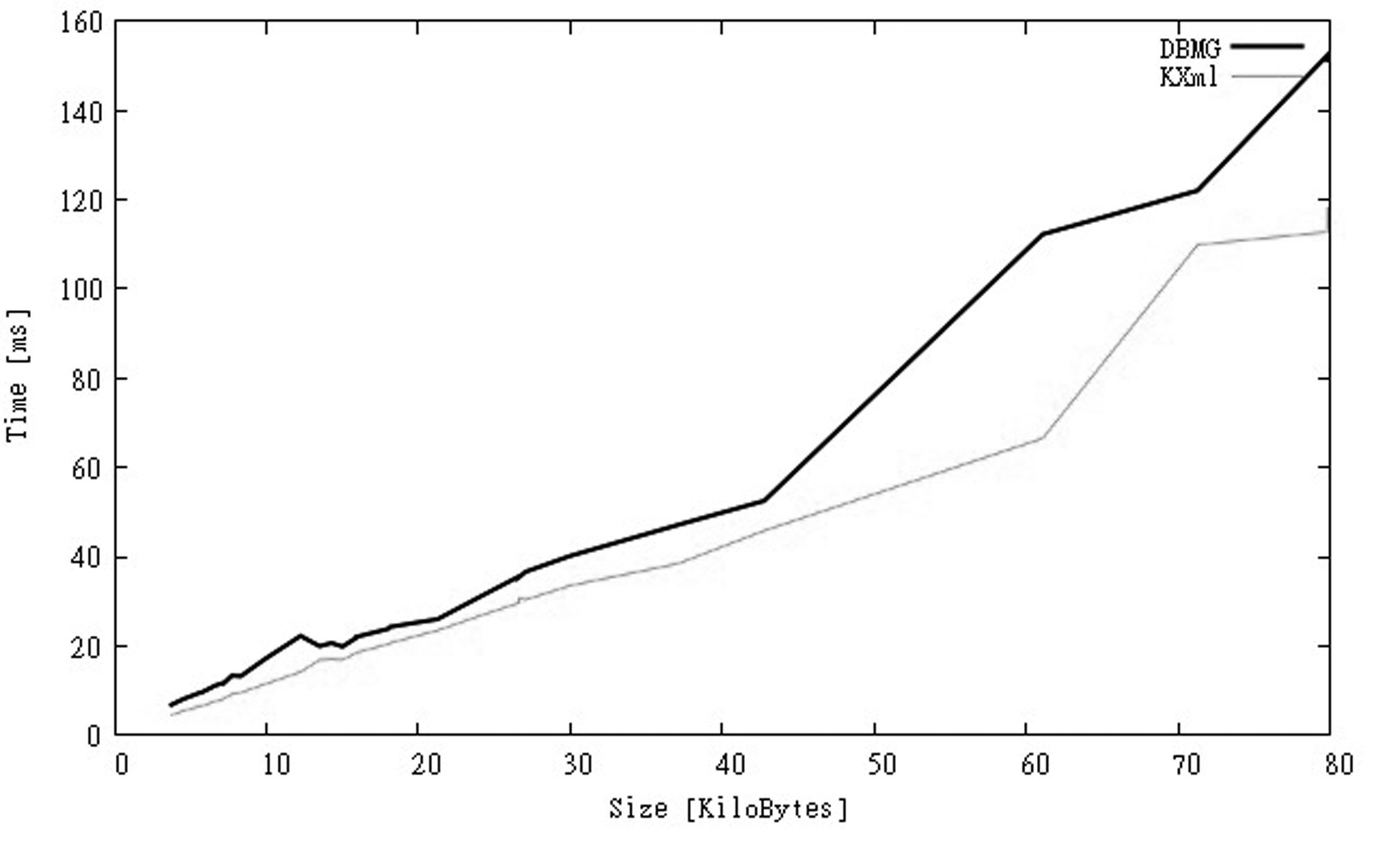}\\
\caption{Execution times for small files}\label{mobile_execution_small}
\end{center}
\end{figure}

\begin{figure}
 \begin{center}
\includegraphics[scale=0.12]{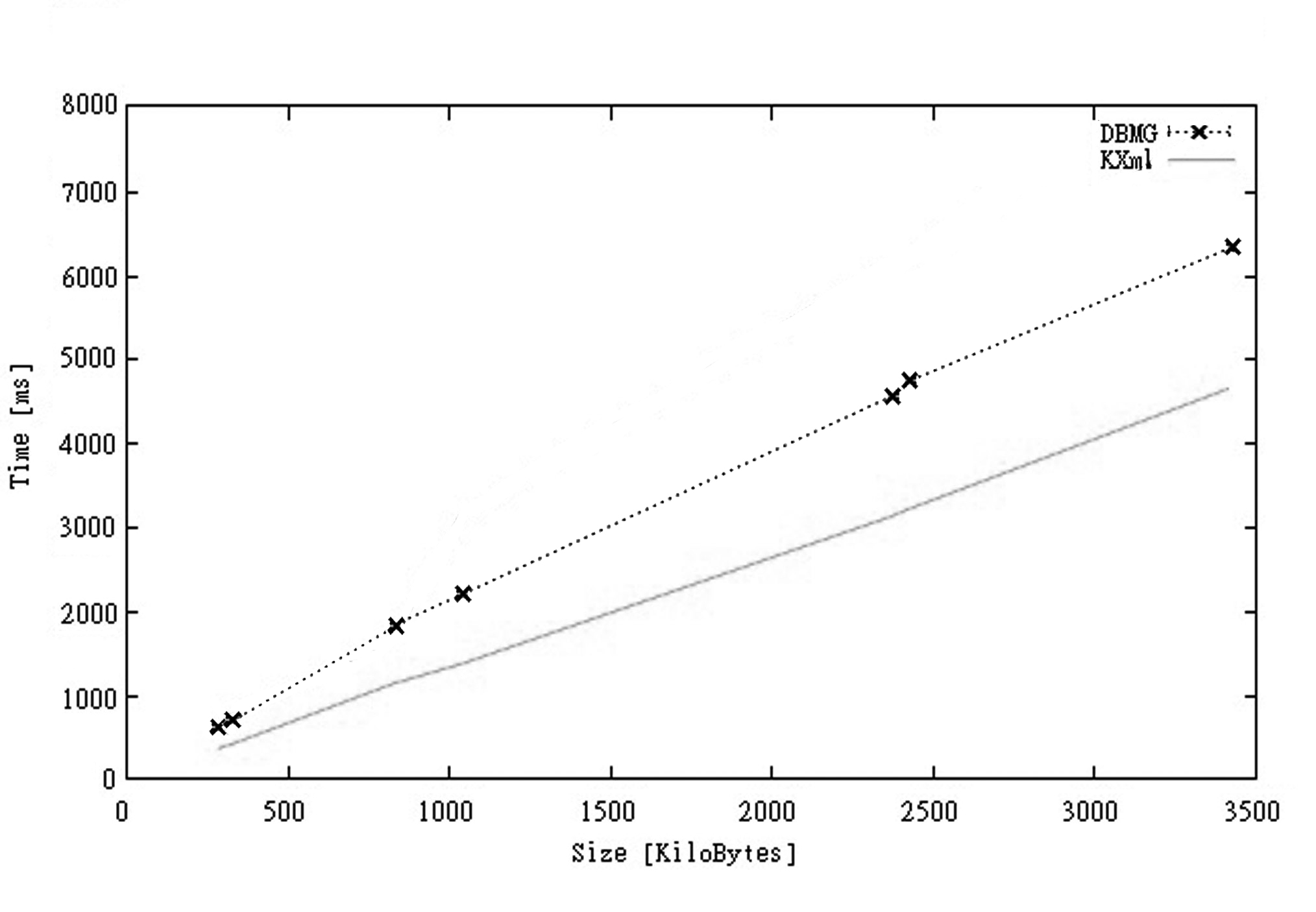}\\
\caption{Execution times for large files}\label{fig:mobile_execution}
\end{center}
\end{figure}

Figures \ref{mobile_execution_small} and \ref{fig:mobile_execution} shows the execution times of code generated by DBMG. The figures also include the execution times needed by \textit{kXML}, the underlying parser used by DBMG, to process the same group of files. The execution times for \textit{kXML} were calculated by creating a simple test case where files are processed using this parser, but no action is taken when receiving the events generated by it.

When files below 100 KB are processed it can be observed that the overhead added by the generated code is not high (Figure \ref{mobile_execution_small}). Nevertheless, we can see in Figure \ref{fig:mobile_execution} that when file size is above 1 MB, the overhead starts to be important (>1s). This happens because the large amount of memory that is required to store the information that is being processed forces the execution of the garbage collector with a high frequency.  We have to keep in mind that code produced manually can have similar problems if  it were necessary to retain most of the information read from the XML files in memory.

The experiment described above was extended in \cite{proc:tamayo6} to compare the code generated by DBMG with other data binding tools and to measure also the performance of this code when executed in a Windows PC. The experiments showed that the execution times for the mobile devices were around 30 to 90 times slower than those for the personal computer. The experiments also showed that the code generated by DBMG was as fast as code generated by other data binding tools for the Android platform.

\section{Conclusions}

In this paper we have presented an approach to generate
compact XML processing code based on large schemas for
mobile devices. It utilises information about how XML documents make use
of its associated schemas to reduce the size of the generated code as much as possible. The solution proposed here is based on the observation that applications that makes use of XML data based on large schemas do not use all of the information included in these schemas. 

A code generator implementing the approach that produces code targeted to Android mobile devices and the Java programming language has been developed.  This tool has been tested in a real case study showing a large reduction in the size of the final XML processing code when compared with other similar tools generating code for mobile, desktop and server environments. Nevertheless, this result must be looked at with caution as the magnitude of the reduction will depend directly from the use that specific applications make of their schemas.

%ACKNOWLEDGMENTS are optional
\section{Acknowledgements}
This work has been partially supported by the ``Espa\~{n}a Virtual'' project (ref. CENIT 2008-1030) through the Instituto Geogr\'{a}fico Nacional (IGN); and project GEOCLOUD, Spanish Ministry of Science and Innovation IPT-430000-2010-11.

%
% The following two commands are all you need in the
% initial runs of your .tex file to
% produce the bibliography for the citations in your paper.

\bibliographystyle{abbrv}

\end{document}